# Two evolutionary lineages: Machiavellian and Bohrian intelligence

**Abstract**

Mutation and natural selection are the two most basic processes of evolution, yet the study of their interplay remains a challenge for theoretical and empirical research. Darwinian evolution favors genotypes with high replication rates, a process called "survival of the fittest" representing lineage of the Machiavellian inteligence. According to quasi-species theory, selection favors the cloud of genotypes, interconnected by mutation, whose average replication rate is highest: mutation acts as a selective agent to shape the entire genome so that is robust with respect to mutation. Thus "survival of the flattest and inventivest" representing lineage of the Bohrian intelligence at high mutation rates. Quasi-species theory predicts that, under appropriate conditions (high mutation pressure), such a mutation can be fixed in an evolving population, despite its lower replication rate.

**Introduction**
The difference in replication rate could be overcome by greater mutational robustness of slower replicator. This robustness was not caused by difference in replication fidelity, but rather by differences in **"canalization"** with respect to mutational perturbation. There is a widespread **trade-off** between intrinsic **replication** rate and **mutational robustness**, which arose during divergence from a common ancestor in environments that differ only in the imposed mutation rate. Thus, mutation rate may depend on particular genotype as well as the environment. The importance of **the mutational cloud** is outgoing from the expectation that **more robust** organisms would **prevail** over faster replicating, but more brittle, organisms at high mutation rate. (2)
During the rebound from the five greatest mass extinctions in Earth's history, it's not an "all-or-nothing" thing. The shape of post-extinction world comes not only from who goes extinct, which of survivors are successful – or, instead, become extinct, or marginalized in the aftermath. Because most extinction event survivor organisms rebound so robustly, many survivors go on lose the evolutionary game. The statistical analysis ruled out one of the most straightforward odd possible causes – that **lineages** that have suffered a major blow to their numbers during a mass extinction might be especially extinction-prone in the aftermath because they contain fewer species to buffer against the hard times.

**Results and Discussion**
**The Machiavellian intelligence:** Gavrilets and Vose´s model base (3) is controversial, but focused primarily on the effects of social factors as the evolution of brain size, climate, and ecology. Their Machiavellian intelligence hypothesis suggests that social power and competition for mates was driving human males to invent strategies that increase cerebral capacity on an evolutionary timescale. Two factors defined intelligence: learning ability (how easily a brain learns new strategies) and cerebral capacity (measures the number of different ideas). These traits get passed on genetically, the more socially potent (intelligent ?) humans winning mates and reproducing more offspring, who inherit their progenitors intelligent genes. It is controversial with recent situation in developmental countries. Gavrilets and Vose described three phases of the dynamics of intelligence: 1. the dormant phase (a default state from which start the evolution), 2. the cognitive explosion phase (increased learning ability and cerebral capacity offer advantages). The cerebral capacity evolves more rapidly than learning ability, suggesting that potential is more important than ability. Even complex ideas are more beneficial for the population, individuals largely gain simple ideas during this phase. The complexity of memes in the population does not increase but, on the contrary, decrease in time. It is a result of intense competition among memes: complex memes give advantage to individuals on a longer (biological) time scale, they lose competition to simpler memes on a shorter (social) time scale. The increasing brain becomes difficult to afford due to high-risk births and energy consumption. The cerebral expansion cannot last forever, and 3. the saturation phase occurs. The natural selection causes a competition between simple and complex ideas. The simple ideas always win because they spread more easily (degeneration). In this phase is promoted a postponing brain growth after birth, and reducing the size of the guts. The complex memes are a selecting factor of the intelligence.

Historically, males with greater social potential ("intelligent fascists" without cosncience) generate more offspring, corresponding to the evolution of more or less (increasing rate of psychopatological skills) intelligent humans. In modern times, intelligence and reproductive fecundity don't generally correlate. The evolution tend to increasingly (and on parallel partly decreasingly) intelligent humans. As the extent to which social success translates into reproductive success declines, cognitive abilities expected to be reduced by natural selection. There are always two evolutionary lineages. The Gavrilets and Vose´s model predicts a reduction in intelligence. The question is remaining: why these large brain dynamics occurred for humans, but not for other animals ? The humans managed to reach a state of ecological dominance was crucial.

According to a study of **the "male warrior effect"**, men need threats, rivalry and war for them to work together the most effectively. The issue of why men start wars is investigated in a series of experiments by M. van Vugt. They reveal that conflict is part of male bonding. Its well known that males are more aggressive than females but with that aggression comes a lot of cooperation too. First of all, male co-operation lies at the heart of democracy and leadership, and men work better in hierarchical groups than women. Men might need wars to show off their altruism, to be celebrated as warriors and heroes. Women leaders are more dovish than hawkish. Results of the study show that rivalry drives males to make sacrifices for their group more than women. This "male warrior effect" causes that men are more likely to support their country going to war, and men are more likely to lead groups in autocratic, militaristic ways. Men have evolved a psychology that makes them particularly interested, and able to engage, in warfare. On the other hand, similar behaviors can be seen in a "pristine primordial form" in chimpanzees. They go out on raids into neighboring communities and **kill off ("survival of the fittest killers")** members of rival groups. Aside from warfare, the males do not usually co-operate.

**The Bohrian intelligence**: Are humans social animals ? Self-sacrifice in social insects only occurs within family groups where genetic interests are strongly overlapping. In contrast, humans have an unrivalled capacity to sacrifice themselves for individuals that are not closely related. Humans invest time and effort in helping the needy within their community, and also make frequent anonymous donations to charities. They come to each other´s aid in natural disasters and respond to appeals to sacrifice themselves for their nation in war time. They put their lives at risk by adding complete strangers in emergency situations. The tendency to benefit others (not closely related) at the expense of oneself is characterizing **altruism** and **prosocial behavior**. This is an important part of the self interest model in group functioning on the basis of the Bohrian intelligence. It is often a complex calculation and ability to put oneself in another person´s shoes or perspective taking, which is a main component of **empathy** and can be viewed as a rule for making a calculated decision (the prefrontal cortex, the lateral intraparietal cortex, the middle temporal area, the superior colliculus). Empathy is really a selfish emotion based on merger between the images of the self and the other, but promotes altruism, increases helping. The experience of a self-conscious emotion, like guilt, should also increase altruism.

The neurobiology of **empathy** (mirror neurons) and **trust** (oxytocin) is helping to understand prosocial behavior within larger groups. Our ability to empathise with others depend on the action of mirror neurons in **the bilateral temporal gyrus** and **the superior temporal sulcus.** Subjects in the several studies who scored higher in empathy tests also showed higher levels of mirror neurons activation. How empathetic we are seems to be related to how strongly our mirror neuron system is activated.

This condition opens up the possibility for **the evolution of social exchange** on the group level, in addition to the interpersonal and intrapersonal level. Group-based altruism (generalized exchange) is beneficial in particularly cooperative situations with highly uncertain pay-offs. Societies also work on the basis of **obedience norms**, (respecting parents), **solidarity norms** (defending their country) what foster group cohesion.

According to G. Edelman, selectional events in the brain are constrained by the activity of **diffuse ascending value systems** (the locus coeruleus, the raphé nucleus, the cholinergic, dopaminergic, and histaminergic nuclei). Their activity affects the selectional process by modulating or altering synaptic thresholds. Value systems also affect systems of learning and memory. The dynamic synaptic changes in individual neuronal groups based on past perceptual categorizations are influenced by limbic and brainstem value systems. The synaptic alterations contribute collectively to a system called a **"value-category memory".** It is based on the activity of frontal, parietal and temporal cortices and is critical to the emergence of consciousness. **Bohrian intelligence** represents a shift to right hemisphere´s conceptual thinking, a more flexible, inventive, creative approach of the metaphorical brain (scientists, economists, writers, physicians, musicians, all with own progressive vision of complex thinking). As M. Fergusson wrote: "The prodigy is devoted to talent, the genius to vision", i. e. to more complex thinking.

**Materials and Methods**

The driving force in evolving cellular life on Earth, has been **horizontal gene transfer**, in which the acquisition of alien cellular components, including genes and proteins, work to promote the evolution of recipient cellular entities. The three main **cellular information processing systems:** translation, transcription and replication suggests that cellular evolution progressed in that order, with **translation** leading the way. The pivotal development in the evolution of modern protein-based cells, was the invention of **symbolic representation** on the molecular level – that is, the capacity to "translate" nucleic acid sequence into amino acid sequence. **Human language** is another example of **the evolutionary potential of symbolic representation.** It has set Homo sapiens entirely apart from primitive relatives, and it is bringing forth a new level of biological organization. Distinct entities during their subsequent evolutions had engaged in **genetic cross-talk** ("jumping genes" of B. Mc Clintock) – they had indulged in a commerce of genes.

S. Gavrilets and A. Vose show that **cerebral capacity** evolves faster and to a larger degree than **learning ability**. Their model suggests that there may by a **tendency** toward a **reduction** in **cognitive abilities** as the reproductive advantage of having a large brain decreases and the exposure to memes (learn strategies) increases in modern societies.

The final limits of the Machiavellian intelligence can be seen from negative tendencies of this type of strategy at the brain size stagnation, climate change, ecology, energy and social problems, etc.

It is now a question, if **the reproductive advantage leading to social power** may be also in 21. century understood as the form of social intelligence. We can see arising global crisis as the result of the Machiavellian intelligence dominance evolved during the last centuries. Behind of decreasing cerebral capacity there could be a shift in intelligence evolution: the Machiavellian intelligence will be changed on the evolutionary scales by the rising **second**

**type** of intelligence: the Bohrian intelligence as **the complementary adaptational lineage** of human development on the global level. The Machiavellian intelligence is declining from its evolutionary progressive role and will be gradually replaced by **the Bohrian intelligence of invention**. As the potential is more important than ability, role of complex ideas will constantly increasing in the future.

The increase of brain becomes difficult due to high-risk births and limits of energy consumptions (accounting only for 2% of the body, the brain uses about 20% of the body´s metabolism). Another big problem with lasting leading role of the Machiavellian intelligence is that historically, males with greater social potency generating more offspring are contributing to evolution of less and less intelligent big amounts of humans in television era. In modern times, the intelligence and reproductive fecundity are correlating inversely (social differences, the deficit of solidarity, big economical crisis, underdevelopment in developmental countries, international crisis, conflicts and wars, etc.).

The extent of social success translated into reproductive success declines in modern societies, and as it showed **Gavrilets and Vose**, the cognitive abilities are expected to be significantly reduced by natural selection based on the strategy of Machiavellian intelligence advantages evolution. We hypothesizing an evolutionary phase shift to the Bohrian intelligence, which could lead to further intelligent evolution of the human race.

Only civilizational phase shift to the Bohrian intelligence can owercome the Le Chatelier´s principle. Because of the deviation from the stady state given by the thermodynamical flow acting against the system is going back to the stady state. The fluctuation of entropy production will be this way positive. There is outgoing a change from dominance (linearity of relations) to **complementarity of two lineages**. The evolutionary goal is to achieve negative fluctuation of the entropy production, to be possible a **new evolutionary step** toward the Bohrian intelligence.

Michael Levin showed with his research (4), that the **electrical fields** help **control cell identity**, cell number, position and movement, which is relevant to everything from embryonic development to regeneration to cancer and almost any biomedical phenomenon you could imagine. In the absence of these fields, cells necessary for regeneration failed to both proliferate and express downstream genetic markers of regeneration. And neuronal growth – long held to be an essential precursor to the generation of other tissues, was disrupted. What we have here is a **master regulator**. And by turning on this one signal, we get **the whole program** of the tail growth.

Probabilities at microscopic level are governed by **interfering probability amplitudes** rather than by additive probabilities. Probability distribution permits us to incorporate wthin the description of the complex microstructure of the phase space. It contains **additional** information that is lacking at the level of linear trajectories. The level of distribution functions permits us to predict the future evolution of the neuronal ensemble.

The Machiavellian intelligence can be mathematically formulated within the framework of **combinatorial-state automata (CSA)** proposed by Chalmers (1994). CSA differ from the finite-state automata (FSA) in that its internal state is not monadic (a lack of internal structure), but has a complex structure (a vector, its elements can be seen as components of overall state, with finite number possible values for each element). These values are the microstates and its combinatorial structure provides the condition for functioning. CSA based on the Machiavellian intelligence can be understood as the mental activity level of **biorobots.** There are many types of biorobots, but the 3 main types are: power maniacs biorobots, sexual biorobots and aggressive (killer) biorobots. (6)

To describe the Bohrian intelligence model we propose t**he Theory of Neuronal Assamblies Coincidence (The TNAC)** hypothesis. The grouping of single neurons into transient assemblies by temporal synchronization is leading to a new level of brain abstractness – metastable states. Synchronized larger abstractness generates the most complex mental states of Bohrian intelligence, including the personal self. Somewhere there is a **shift from cognitive to conceptual** operations of thinking. **The grouping based on the coincidence of the neural assemblies** is the critical step the binding process. The neurons in both cortical and subcortical levels can synchronize their activity in the millisecond range. Such synchronization tunes preferentially to a particular features and this process relay on self-organization. The level of organization at which cognition, thinking and consciousness operates is a highly organized macro-level of electrophysiological phenomenon in the brain, is programmed by the coordinated electrical activity in the neuronal populations (4). We can propose that the information carried by the system results due a **selection synchronized by coincidence (probability amplitudes overlap)** from the molecular to a subatomic level from many differing possibilities (article in preparation). The problem with Machiavellian intelligence is now clear, because of its counterproductive and partly destructive character on the long term. Decreasing efficacy of the Machiavellian intelligence is causing on the evolutionary level is replacing with a second Darwinian lineage: higher adaptability of the Bohrian intelligence (can be described by the TNAC), based on the advantages of robust mutations ("survival of the flattest but inventivest").

According to **the costly signaling theory**, a well-established framework in biology and economics, may be useful to analyze the individual differences in human unconditional altruism. We propose that unconditional altruistic behavior is related to general intelligence. The cost of unconditional **altruism** is lower for **highly intelligent** people than for less intelligent people because they may expect to **regain** the drained resources. As result, **unconditional altruism** can serve as an honest **signal of intelligence**. Distinguishing altruistic behavior from cooperative behavior in social psychological and economic theories of human behavior might be useful.

In this process **trust** plays a key role in **organizations, economic exchange** and **politics.** The functioning **market economy** requires trust because if you do not trust in your business partners, **market transactions break down**. Also **functioning democracy** requires trust because if you lose trust in a **country´s leaders** and **institutions**, **political legitimacy break down.** It would be possible to enhance trust if we will know more on neurobiological mechanisms of ability to trust during for example business transactions. The hormone **oxytocin** plays a key role in human trust and specifically increases the individual´s willingness to **accept social risks**. (7) Oxytocin is a part of the biological basis of **prosocial and profuture approach behavior.**

ACKNOWLEDGEMENTS. I am particularly grateful to Vladimir Varga, Konrad Balla and Livuska Ballova for discussions and comments.

**Robert Skopec**
**941 35 Dubník 299**
**Slovakia**

**E-mail: zxcbnvm@yahoo.co.nz**

20.5.2007
R. Skopec